# Growth-rate dependent partitioning of RNA polymerases in bacteria

Stefan Klumpp[*] and Terence Hwa

*Center for Theoretical Biological Physics and Department of Physics,*

*University of California at San Diego, La Jolla, CA 92093-0374*

[*]To whom correspondence should be addressed. *e-mail: klumpp@ctbp.ucsd.edu*

**Corresponding Author:**
Dr. Stefan Klumpp,
Center for Theoretical Biological Physics,
University of California at San Diego,
9500 Gilman Drive, La Jolla, CA 92093-0374
phone: 858 534-7256; fax: 858-534-7697;
email: klumpp@ctbp.ucsd.edu

# Abstract

Physiological changes which result in changes in bacterial gene expression are often accompanied by changes in the growth rate for fast adapting enteric bacteria. Since the availability of RNA polymerase (RNAP) in cells is dependent on the growth rate, transcriptional control involves not only the regulation of promoters, but also depends on the available (or free) RNAP concentration which is difficult to quantify directly. Here we develop a simple physical model describing the partitioning of cellular RNAP into different classes: RNAPs transcribing mRNA and ribosomal RNA (rRNA), RNAPs non-specifically bound to DNA, free RNAP, and immature RNAP. Available experimental data for *E. coli* allow us to determine the two unknown parameters of the model and hence deduce the free RNAP concentration at different growth rates. The results allow us to predict the growth-rate dependence of the activities of constitutive (unregulated) promoters, and to disentangle the growth-rate dependent regulation of promoters (e.g., the promoters of rRNA operons) from changes in transcription due to changes in the free RNAP concentration at different growth rates. Our model can quantitatively account for the observed changes in gene expression patterns in mutant *E. coli* strains with altered levels of RNAP expression without invoking additional parameters. Applying our model to the case of the stringent response following amino acid starvation, we can evaluate the plausibility of various scenarios of passive transcriptional control proposed to account for the observed changes in the expression of rRNA and biosynthetic operons.

\body

# INTRODUCTION

Bacteria are able to grow with wildly different growth rates in different media. Depending on the growth conditions, the quality and availability of nutrients, they differ in cell size and macromolecular compositions, e.g., the ratio of protein, RNA, and DNA (1, 2). For bacteria in exponential growth phase, this dependence was found empirically as a dependence on *growth rate* rather than as a dependence on the specific growth medium, since bacteria grown in different media that support the same growth rate exhibited the same macromolecular composition (1-3). For this reason, many parameters of the bacterial cell have been characterized as functions of the growth rate (4). Many of these parameters affect gene expression, e.g., the cellular abundance of transcription and translation machinery. Gene expression is therefore expected to exhibit a generic growth-rate dependence in addition to the specific genetic regulation (5). Indeed, even unregulated (or "constitutively expressed") promoters exhibit growth-rate dependent activities (5, 6). Some genes, e.g., the ribosomal RNA operons (*rrn*), are additionally regulated in a growth-rate dependent fashion (7, 8).

One difficulty in elucidating various mechanisms of growth-rate dependent transcriptional control lies in the fact that the activity of a promoter depends not only on the active control mechanisms, but also directly on the availability of RNA polymerase (RNAP) which is growth-rate dependent. For example, the total number of RNAPs per cell was determined to increase from 1500 at slow growth (0.6 doublings/hour) to 11400 at fast growth (2.5 doublings/hour) (4). How the concentration of *free* RNAPs, which is crucial to the initiation of transcription, depends on growth rate is less clear. Nevertheless, "passive transcriptional control" (3), i.e., changes in gene expression due to changes of the free RNAP concentration alone, was proposed to play a role in the growth-rate dependent regulation of rRNA transcription (7, 9), based on observations that similar behaviors could be induced by RNAP mutations (9, 10). Passive control has also been proposed to account for changes in transcription upon sudden depletion of nutrients, during the so-called "stringent response". Surprisingly, both decreasing and increasing free RNAP concentrations have been proposed to occur during the stringent response, and were invoked by different authors to explain either the downregulation of *rrn* operons (6, 9) or the upregulation of biosynthetic operons (10, 11). These proposals are hard to test experimentally, as the concentration of the free RNAPs in cells is difficult to measure directly. Also, indirect inference based on measurements of the cytoplasmic fraction of RNAPs (12, 13) and promoter activities (6, 14) rely on assumptions that may be questioned (see below).

In this study, we developed a method to estimate the free RNAP concentration in *E. coli* cells growing with different growth rates. Our method is based on a physical model which partitions the RNAPs in a cell into fractions representing RNAPs transcribing mRNA and rRNA, RNAPs non-specifically bound to DNA, free RNAPs, and RNAP assembly intermediates. Our model combined features from previous studies of RNAP partitioning (15-17), none of which however included all these fractions. By integrating the available data from both direct and indirect measurements of the free RNAP concentration together with the growth-rate dependence of the macromolecular composition of *E. coli* cells (4), this model allowed us to predict the growth-rate dependent partitioning of RNAPs, thereby providing a quantitative picture of the various activities of RNAPs in the cell. The results for the concentration of free RNAP allowed us to predict the growth-rate dependence of the activities of the constitutive promoters, as well as to disentangle the various growth-rate dependent factors affecting the activity of the *rrn* promoters. We finally applied our model to investigate the change in free RNAP concentration during the stringent response and test several scenarios for passive control. The results suggest that passive control, both positive and negative, should not be expected to play a major role in the stringent response, at least in the early stage immediately following sudden starvation.

# MODEL AND RESULTS

The concentration of free RNAPs in cells is difficult to measure. Two approaches have been described in the literature. The first one is indirect and uses transcription from a constitutive (unregulated) promoter (6, 7, 14, 18). This approach yields only RNAP concentrations relative to the Michaelis constant of that promoter. To estimate an absolute value of the free RNAP concentration under this approach, one has to rely on the kinetic parameters of the promoter measured *in vitro* (18), which depend on experimental conditions and may not be representative for the situation *in vivo*. The interpretation of such data is further complicated by controversies about whether specific promoters [in particular, the rRNA promoter P2 used in refs. (6, 7)] are actually constitutive; see below. The second, more direct approach is to use DNA-free mini-cells and compare the RNAP content of mini-cells and normal cells to obtain the fraction of cytoplasmic RNAPs (12, 13). The cytoplasmic RNAP measured in these experiments however includes free RNAPs as well as RNAP assembly intermediates, and possibly also other forms such as RNAPs sequestered on 6S RNA (19) in stationary or very slowly growing cells. The advantage of this approach is that it yields absolute RNAP concentrations. However, to link the results to the free RNAP concentration, we need to understand quantitatively the partitioning of total RNAP, which is the subject of this section.

**Model for the partitioning of RNA polymerases**

We developed a model for the partitioning of RNAP based on the assumption that in exponentially growing cells, all RNAP in the cell fall into one of five different classes: (i) RNAPs transcribing mRNA, (ii) RNAPs transcribing rRNA, (iii) RNAPs non-specifically bound to DNA, (iv) free RNAPs in the cytoplasm available for transcription, and (v) RNAP subunits and assembly intermediates. Some of these classes require further explanation.

(a) Non-specific binding of RNAP to DNA, as demonstrated *in vitro* (20, 21), is much weaker than the specific binding of RNAPs to promoters, but is nevertheless expected to play an important role *in vivo*, because the number of sites for non-specific binding greatly exceeds the number of promoters (22, 23). *In vitro*, non-specifically bound RNAPs have been directly shown to slide along DNA (24, 25), which may play a role in the kinetics of promoter binding. *In vivo*, non-specific binding has not been directly demonstrated for RNAP. Non-specific binding (26) and sliding along DNA (27) have however been demonstrated *in vivo* for transcription factors, which exhibit non-specific binding to DNA similar to that of RNAPs *in vitro* (22, 28). Furthermore, non-specific binding of RNAP *in vivo* is consistent with the observation that a large fraction of RNAPs, larger than the fraction of actively transcribing RNAPs, is associated with the nucleoid (13).

(b) Intermediates of RNAP assembly (immature RNAPs) have to be taken into account (13), because the RNAP content of cells is determined by measuring the fraction of total protein mass that is in β and β' subunits of RNAP (4). Some of these subunits are not or are only partially assembled into functional RNAPs in the cell (29). These assembly intermediates are located in the cytoplasm and radioactively labeled β and β' subunits appear in the nucleoid after about 5 and 2.5 minutes, respectively (29). The larger of these times is likely to correspond to the time needed to fully assemble an RNAP plus the transition to the nucleoid, thereby providing an upper bound for the maturation time $\tau$.

(c) Under conditions different from exponential growth there may be fractions of RNAPs in addition to the five listed above. Additional classes are clearly present under various stress conditions and for cells in the stationary phase, where alternative sigma factors play important roles, so that RNAPs have to be partitioned according to their sigma factors (16, 30), and where a fraction of RNAPs is inactivated by the regulatory 6S RNA (19). In this study, we focus on exponentially growing cells, for which the concentration of the housekeeping sigma factor, sigma 70, is very high (16, 31), and concentrations of alternative sigma factors are considerably lower (16, 30). In addition, alternative sigma factors have lower affinities for core RNAP than sigma 70 (32). These two features allow us to neglect the competition of sigma factors for exponentially growing cells. Furthermore the affinity of sigma 70 for the RNAP core enzyme is very high (32, 33), so that essentially all free RNAPs are bound to sigma factor.

(d) The models of Bremer et al. (15) and of Tadmor & Tlustly (17) consider RNAPs pausing in transcription as an additional class. We take pausing to be an integral part of transcript elongation, as the measured elongation speeds are average values that include pauses (34, 35). An incentive for Bremer et al. to separate pauses from active transcription is the assumption that there are specific "pause genes", for which pausing is strongly enhanced during the stringent response, so that these genes sequester RNAPs. There is however little experimental support for the existence of such pause genes and the transcription speeds assumed in these models are far lower than measured values (34, 35). The main difference between our model and the models of Refs. (15, 17) is thus the description of non-transcribing RNAPs associated with the nucleoid: In our model, these RNAPs are considered as non-specifically bound to DNA, while non-specific binding is not included in the models of Refs. (15, 17), where these RNAPs are assumed to be pausing in transcription.

To determine the partitioning of RNAPs into these five classes, we derived quantitative expressions for the numbers of RNAPs in each class that link these numbers to measured parameters of the cell (Fig. 1 and Supporting Text). The numbers of RNAPs transcribing mRNA and rRNA ($N_m$ and $N_r$) are estimated directly from measured RNA synthesis rates ($r_m$ and $r_r$) and RNAP speeds ($c_m$ and $c_r$) at different growth rates. In addition we link these numbers to the biophysical properties of the corresponding promoters using a Michaelis-Menten model of transcription activity, which relates the transcription rates to the concentration $c_{\text{free}}$ of free RNAP (15). This description is used below to study growth-rate dependent regulation of transcription.

The main task of our model is to quantify the partitioning of the non-transcribing RNAPs into the other three classes, namely free RNAPs ($N_{\text{free}}$), non-specifically bound RNAPs ($N_{\text{ns}}$) and assembly intermediates ($N_{\text{interm}}$). In our model, the number of non-specifically bound RNAPs is determined by equilibrium binding to DNA, with a growth-rate dependent number of possible binding sites. As mentioned above, sliding of non-specifically bound RNAPs along DNA may play a role for the *kinetics* of promoter binding; this kinetic effect is not explicitly described in our model, which describes only the (quasi-)equilibrium binding of RNAPs to promoters and non-specific sites. In this thermodynamic description, the numbers of transcribing RNAPs depend only on the concentration of free RNAPs even though non-specifically bound RNAPs may start transcribing without dissociation from the DNA. The number of free RNAPs is described by a concentration of free RNAPs ($c_{\text{free}}$), which we determine below as a function of growth rate, and the cellular volume ($V_C$), and the number of immature RNAPs is described by a maturation time $\tau$ after which newly synthesized RNAPs are functional.

Most parameters of our model have been measured [see ref. (4), Supporting Text and Supporting Tables S1 and S2], but the model contains two unknown parameters, the dissociation constant for non-specific RNAP-DNA binding, $K_{\text{ns}}$, and the maturation time $\tau$ of newly synthesized RNAPs. We assume these parameters to be independent of the growth rate themselves and determine them by matching the fraction of cytoplasmic RNAPs predicted by our model to data from mini-cell experiments (12, 13), see Supporting Text and Figure S1. This procedure leads to a maturation time $\tau$ of 3.4 min and a dissociation constant for non-specific binding of 3100 μM. These values are consistent with experimental data, as discussed in Supporting Text, and will be used throughout the following. However, two-fold changes in the values of these two parameters lead to very similar results (Figure S2). We note that the dissociation constant might be growth-rate dependent if the level of macromolecular crowding changes with the growth rates (17). This question has not been addressed directly by experiments, but indirect evidence does not suggest a strong change; see Supporting Text.

**Predicted growth-rate dependence of RNAP partitioning**

Using the above model, we computed the partitioning of RNAP into each of the 5 classes for growth rates ranging from 0.6 to 2.5 doubling/hour; the results are shown as total number per cell in Fig. 2A and as concentration (after taking into account the growth-rate dependent cell size, see Table S1) in Fig. 2B. In these plots, grey symbols indicate the species of transcribing RNAPs that are estimated directly from RNA synthesis rates (using Eqs. [1b] and [2b] in Fig. 1), while colored symbols indicate the predicted partitioning of the non-transcribing RNAPs: non-specifically bound (blue), free (red), assembly intermediates (green). Fig. 2A shows that the actual numbers of RNAPs per cell (measured and predicted) increase with the growth rate for each of the five species. However the numbers of RNAPs transcribing rRNA (grey circles) and those involved in assembly (green triangles) increase more strongly (44-fold and 31-fold, respectively) than the numbers of RNAPs of the other species (at most 9-fold) and also more strongly than the measured total number of RNAPs per cell (black circles), which exhibits a 7.6-fold increase.

Fig. S3 shows the same RNAP partitioning as fractions of the total RNAP number. For all growth rates, non-specific binding to DNA (blue triangles) is predicted to make up the largest fraction of RNAPs, despite the fact that non-specific binding is very weak. Non-specific binding accounts for 75 percent of all RNAPs at slow growth (0.6 doubling/hour). This fraction decreases to 54 percent at 2.5 doublings/hour. The strongest increase is seen for the fraction of RNAPs transcribing rRNA, which increases about 5.8-fold, from 4 percent to 23 percent. Likewise the fraction of RNAP assembly intermediates exhibits a 4-fold increase, while the fraction transcribing mRNA and the fraction of free RNAPs exhibit only small changes (less than two-fold). Note that the total fraction of cytoplasmic RNAP (red + green) is below 20% and the fraction of assembly intermediate is below 10% even at the highest growth rate studied.

We finally turn to the free RNAPs which is the focus of our study. As shown by the red curve in Fig. 2B, the concentration of free RNAPs is predicted to increase from 0.47 µM for a growth rate of 0.6 doublings per hour to 1.1 µM for 2.5 doubling per hour. The range of the free RNAP concentration is substantially higher than the estimate of 30 nM made by McClure (for a doubling time of 50-60 min) based on a comparison of *in vivo* transcription rates from various promoters with their Michaelis constants measured *in vitro* (18). Our result (red curve of Fig. 2B) has two remarkable features which will be elaborated below: (i) the overall change of the free RNAP concentration over the studied range of growth rates is only about two-fold, significantly less than previous estimates (6, 14, 15), and (ii) the growth rate dependence of the free RNAP concentration *saturates* at high growth rates.

## DISCUSSION

**Constitutive promoters**

The transcription from unregulated (constitutive) promoters is expected to depend on the growth rate in a way that is completely determined by the growth-rate dependence of the free RNAP concentration.[1] Based on this idea, Liang et al. have studied the transcription of promoters believed to be constitutive in order to determine the growth rate dependence of the free RNAP concentration (6). They found that, at slow growth rates, the transcription from these promoters[2] increased approximately in parallel with increasing growth rate, i.e. the ratio of the levels of transcription from these promoters remained approximately constant (see also Fig. S4 A and B). At fast growth, transcription rate from most promoters saturated, but transcription from the ribosomal RNA promoter P2 kept increasing [purple line in Fig. S4A or Ref. (6)]. Liang et al. suggest the following interpretation of these results: The increase of transcription of *rrn* P2 reflects the increase of the free RNAP concentration, as P2 appears not to be saturated with RNAPs under their experimental conditions. The other constitutive promoters however become saturated with RNAP at fast growth, thereby reflecting the increase of the free RNAP concentration only at slow growth.

Although the argument of Liang et al. is very elegant, this interpretation has several difficulties. First, the parallel increase of transcription from these promoters is only approximate. Comparing the two smallest growth rates studied by Liang et al. (where all their promoters should be far from saturation with RNAPs), the increase in transcription varies between 1.4-fold and 2-fold (see also Fig. 3). This may not be sufficient to distinguish constitutive expression from weakly regulated expression. Second, a comparison of their results for wild type cells and for the relaxed strain devoid of ppGpp shows that, at a given growth rate, the transcription from P2 is almost the same in both strains [Fig. 3 of ref. (6), see also Figure S5A]. According to their interpretation, the free RNAP concentration is thus also the same in both strains and one would expect the transcription rates of other constitutive promoters also to be the same. Their data show however that the transcription from most other constitutive promoters is reduced in the relaxed strain at high growth rates compared to the wild type [Fig. 2 of ref. (6), see also Figure S5B]. Finally, the method is based on the assumption that P2 is a constitutive promoter, which is controversial (see Supporting Text).

---

[1] This assumes again that there is no strong effect due to changes in macromolecular crowding, see Supporting Text.
[2] In this study, transcription rates were determined by measuring the beta-galactosidase activity for LacZ expressed from the promoter of interest. The relative beta-galatosidase activity obtained from two different promoters *at the same growth rate* provides a measure of the relative transcription rate. Absolute values of the transcription rates have been determined from the relative activities of the promoters compared to that of the rrn promoter pair P1-P2 and the absolute values of the transcription rate from P1-P2 as obtained from the rRNA content of the cells and the *rrn* operon multiplicity (5).

Using our result for the free RNAP concentration, we can predict the growth-rate dependence of the transcription rate for an unregulated promoter, i.e. the rate of mRNA synthesis, which corresponds to the mRNA level assuming that mRNA lifetime is not growth-rate dependent. The growth-rate dependence of an unsaturated constitutive promoter should be proportional to the growth-rate dependence of the free RNAP concentration. In Fig. 3, we plotted the data of Liang et al. (6) for the growth rate dependence of several constitutive promoters ($P_{bla}$, $P_{spc}$, P2) together with the growth-rate dependence of the free RNAP concentration (red curve of Fig. 2B). All curves were normalized to their respective maximal value. This plot shows that the growth-rate dependence of those promoters with saturating expression at fast growth ($P_{bla}$, $P_{spc}$) approximately parallels the growth-rate dependence of the free RNAP concentration. This observation suggests a rather different interpretation of the data of Liang et al.: Transcription from these promoters directly reflects the free RNAP concentration at *all* growth rates. The apparent saturation of these promoters at fast growth does not indicate that these promoters are saturated with RNAPs, but according to our picture, results from the fact that free RNAP saturates for high growth rates. This interpretation also suggests that the observed increase of transcription from the P2 promoter of rRNA (6, 14) is due to growth-rate dependent regulation, with the implication that P2 is not a constitutive promoter. (See Supporting Text for a detailed discussion of the P2 promoter, including a review of the salient arguments in the literature.) We note that the two interpretations could be distinguished experimentally by overexpressing RNAP in fast growing cells. While the interpretation suggested here predicts an increase in the transcription of the constitutive promoters, the original interpretation of Liang et al. predicts that transcription from the constitutive promoters should be unaffected by the increased RNAP level. Such an experiment has so far only been done with slowly growing cells (see below).

**Growth-rate dependent regulation of promoter activity**

Next, we used our result for the free RNAP concentration to study the growth-rate dependent regulation of the *rrn* promoters. (A corresponding calculation for the average mRNA-synthesizing promoter is described in Supporting Text.) Over the range of growth rates studied here, there is a 44-fold increase in rRNA synthesis (Table S1), which is based on a 2.9-fold increase in of the operons copy number (4) and a 2.3-fold increase of the free RNAP concentration (Fig. 3). As a consequence, a 6.6-fold increase in *rrn* transcription is achieved by an increase in promoter strength, which reflects growth-rate dependent regulation, isolated from the change in free RNAP concentration.[3] To determine promoter strengths at different growth rates, we use a Michaelis-Menten model of transcription (Eqs. [1a] and [2a]), with the ratio of the maximal transcription rate and its Michaelis constant, $A_r=V_r/K_r$ taken to be a measure of the promoter strength. Fig. 4A shows the promoter strengths for the promoter pair P1-P2 of the *rrn* operons as well as those for the individual *rrn* promoters P1 and P2, calculated in this way from the transcription rates measured in Ref. (14). The P1-P2 promoter strength $A_r$ increases about 80-fold over the studied range of growth rates [calculating $A_r$ from the transcription rates given in ref. (4) leads to smaller values, but a similar growth-rate dependence, see Fig. S6A]. The P1 promoter exhibits strong growth-rate dependent regulation, with a predicted ~1000-fold increase of its promoter strength. As discussed above, we expect the strength of P2 to be regulated as well, but Fig. 4A suggests that its regulation is much weaker than that of P1, with only a 5.4-fold increase of the promoter strength. We note that over the range of growth rates where Liang et al. observed co-variation of P2 with constitutive promoters (~0.6 -1.3 doublings-hour) (6), our model gives only a ~1.5-fold change in the strength of P2, which is probably too small to be distinguished from co-variation in that experiment. Our conclusion of a weak growth-rate dependent regulation of P2 is in agreement with the conclusion of Murray et al. (36) based on *in vitro* studies; see also Supporting Text. But our model additionally allows us to separate the regulation of the promoter from the growth-rate dependence of the free RNAP concentration.

**Transcription with over- and under-production of RNAP**

---

[3] The increase in promoter strength is expected to be larger than 6.6-fold, because the promoter approaches saturation with RNAPs for fast growth (see Supporting Text for estimates of the maximal transcription rate).

We next used our model to study the effect of changing the total amount of RNAPs per cell. This has been studied experimentally by Nomura et al., who either increased the level of RNAPs per cell by expressing RNAP core enzyme subunits on a plasmid or decreased the RNAP level by replacing the chromosomal β and β' genes with β and β' genes controlled by the *lac* promoter and controlling its level of induction (37). For *E. coli* growing on glycerol-amino acids medium (with a growth rate of ~1.5 doublings/hour), they found that an up to two-fold change of the amount of RNAP per cell in either direction had little or no effect on the growth rate and on the transcription of rRNA, but resulted in a proportional change in the transcription of both the total mRNA and of the mdh mRNA, with the latter serving as a probe for the transcription from an unregulated promoter (37). To check whether our model can account for this result, we varied the total RNAP number and determined the predicted transcription rate of mRNA (see Supporting Text). The results shown in Fig. 4B (black bars) are in excellent agreement with the data of Nomura et al. (37) without invoking any additional parameters.

**Passive control in the stringent response**

Finally, we addressed the change in free RNAP concentration during the stringent response and used our model to test several scenarios for the passive control of rRNA or mRNA synthesis. As mentioned in the Introduction, both increasing and decreasing free RNAP concentrations have been proposed during the stringent response. We consider the immediate response to starvation (within the first few minutes), before the composition of the cell, in particular the RNAP and ribosome content, is substantially changed. This situation is implemented in our model by changing one or several parameters according to what was measured during the stringent response, while keeping all other parameters fixed at the values they had before starvation. This simplification is based on the fact that the parameter changes are due to an increase of the cellular concentration of the regulatory nucleotide ppGpp, which increases very quickly (38, 39), while changes in protein content are expected to occur more slowly. We then calculate the partitioning of RNAPs according to the changed parameters.

We first tested the proposal that a decreased mRNA elongation speed ($c_m$) sequesters RNAPs in transcription and that rRNA synthesis is downregulated by the resulting reduction of the free RNAP concentration (6, 9). The elongation speed of mRNA is reduced to 19-28 nt/s during the stringent response, most likely due to increased RNAP pausing induced by ppGpp (38, 40). The elongation of rRNA is unchanged at their value of 85 nt/s, since RNAPs transcribing rRNA are protected against ppGpp-induced pausing by the rRNA antitermination complex (40). When we changed $c_m$ from 55 nt/s to 20 nt/s, our model exhibits a decrease of the free RNAP concentration of only 13 percent (Fig. 4C, white bar). [The exact value depends on the growth-rate before starvation, which is 2.5 doublings per hour in Fig. 4C; the results for slower growth rates are very similar, see Fig. S7 A]. This decrease of the RNAP concentration has a very small effect on the synthesis of rRNA (<10 percent, Fig. S7 B), while measured reductions are at least 5-10-fold, see e.g. refs. (39, 41, 42). Our model predicts that much slower $c_m$ would be needed to affect rRNA synthesis substantially. For example, to obtain even a modest 2-fold suppression of rRNA synthesis by sequestering RNAPs in transcription, we have to reduce $c_m$ to below 5 nt/s (data not shown), which is far below the experimentally observed range. We can thus conclude that sequestering of RNAPs in transcript elongation plays only a minor role in the suppression of rRNA synthesis during the stringent response. This conclusion is in agreement with experimental results for a NusA mutant, which exhibits a normal stringent response without a reduction of the mRNA elongation speed (41).

We then tested whether the suppression of rRNA synthesis due to ppGpp-dependent regulation of the *rrn* promoters (7, 8) increases the free RNAP concentration. This increase has been proposed to explain the positive regulation of biosynthetic operons during the stringent response (10, 11). We find however that, even if rRNA synthesis is shut off entirely, the free RNAP concentration increases only by 35 percent (Fig. 4C, third bar). A more realistic estimate for the suppression of rRNA synthesis ($K_r \approx 20$ μM during the stringent response, see Supporting Text) leads to almost the same result (fourth bar in Fig. 4C). A further increase of the free RNAP concentration could be due to the repression of a fraction of protein-coding operons (other than the biosynthetic operons). However, even for the extreme case that all transcription (mRNA and rRNA) is completely stopped in the stringent response, we obtained only a 1.5-fold increase of the free RNAP concentration (Fig. 4C, fifth bar), not sufficient to explain the observed 2-3-fold stimulation in the transcription of, e.g., the *his* operon (43). We finally studied the combined effect of a decrease in mRNA elongation speed and decreased transcription of rRNA and found a weak decrease of the free RNAP concentration (by ~10 percent) for slowly growing cells (violet curve, Fig. S7) and a similarly weak increase (up to 15 percent) for fast growing cells (Fig. 4C, bar 6). These results strongly suggest that passive control by increased RNAP concentration plays only a limited role in the positive control of biosynthetic operons during the stringent response. This conclusion is consistent with the recent experimental demonstration that ppGpp has a direct stimulating effect on the transcription of biosynthetic operons *in vitro*, an effect not noticed before because it requires the co-regulator DksA (43). The relative importance of direct and passive effects has however remained unclear and our results suggest that the direct control dominates.

Finally, evidence has accumulated in recent years indicating that altered sigma factor competition plays an important role in the stringent response (44-47). This is to some extent an effect of altered expression of sigma factors and their regulators such as anti-sigma factors (42) and therefore expected to be important during later stages rather than in the immediate response. However direct effects of ppGpp on sigma factors, e.g. favoring at least some alternative RNAP-sigma complexes, may also contribute. It has therefore been proposed that during the stringent response, the concentration of free RNAP with bound sigma 70 is reduced (45). An alternative proposal suggests that due to the suppression of rRNA transcription, more RNAP core enzymes become available to bind alternative sigma factors, so that operons controlled by alternative sigma factors could be up-regulated passively (46, 47). Based on our analysis above we expect the latter effect to be small; but at the moment we cannot test these ideas quantitatively, because the effect of ppGpp on the formation and the activity of alternative holoenzymes is unclear and important parameters such as affinities of sigma factors to core RNAP and sigma factors concentrations are unknown in the stringent response. This important question must therefore be postponed to future research.

**Acknowledgments**

We are grateful to Eddy Mateescu, Hans Bremer, Pat Dennis, Rick Gourse, Sarah Ades, and Elio Schaechter for discussions during the course of this work. This work was supported by Deutsche Forschungsgemeinschaft (fellowship KL818/1-1 and 1-2 to SK) and the National Science Foundation through the Center for Theoretical Biological Physics (grants PHY-0216576 and 0225630), as well as through NSF grant MCB 0746581 (to TH).

**FIGURE LEGENDS**

**Figure 1: Model for the partitioning of RNAP.** In exponentially growing cells all RNAPs are taken to fall into one of five classes, RNAPs transcribing mRNA ($N_m$) and rRNA ($N_r$), RNAPs non-specifically bound to DNA ($N_{ns}$), free RNAPs ($N_{free}$), and RNAP assembly intermediates (immature RNAPs, $N_{interm}$). The total number of RNAPs per cell ($N_{RNAP}$) is the sum of the number of RNAPs in these classes. Our model describes the numbers of RNAPs in each class by equations that link them to measured biophysical parameters of the cell (see Supporting Text for a detailed description and Supporting Tables S1 and S2 for the parameter values, many of which are growth-rate dependent). The numbers of transcribing RNAPs ($N_r$ and $N_m$) are both described by a microscopic model Eqs. [1a], [2a] and estimated directly from measured RNA synthesis rates Eqs. [1b] and [2b].

**Figure 2: Partitioning of RNAPs at different growth rates.** (A) Total number of RNAPs per cell and numbers of RNAPs in the different classes as predicted by our model. (B) Concentrations of total RNAP and RNAPs in the different classes.

**Figure 3: Growth-rate dependent transcription from constitutive promoters.** Growth-rate dependence of the transcription rates from several constitutive promoters and the *rrn* promoter P2. Data are taken from ref. (6) and have been normalized to the maximal value per promoter. [for P2 we also included corresponding data from ref. (14).] The black curve indicates the free RNAP concentration from Fig. 2, which is proportional to the predicted transcription rate from an unsaturated constitutive promoter.

**Figure 4: Consequences of the predicted free RNAP concentration.** (A) Growth-rate dependent regulation of the *rrn* promoters: Effective promoter strengths for the *rrn* promoters P1 (black), P2 (grey), and the pair P1-P2 (white) as calculated from the transcription rates measured in ref. (14). (B) Predicted mRNA expression for over- and under- expression of RNAP and comparison to data from ref. (37). (C) Passive control during the stringent response: Concentration of free RNAPs during the stringent response relative to the concentration during the exponential growth (with a rate of 2.5 doublings/hour) before starvation.

$$N_{\text{RNAP}} = N_{\text{m}} + N_{\text{r}} + N_{\text{ns}} + N_{\text{free}} + N_{\text{interm}}$$

**Non-transcribing RNAPs:**

$$N_{\text{ns}} = gG_{\text{C}} \frac{c_{\text{free}}}{c_{\text{free}} + K_{\text{ns}}} \quad [3]$$

$$N_{\text{free}} = c_{\text{free}} V_C \quad [4]$$

$$N_{\text{interm}} = N_{\text{RNAP}}(1 - 2^{-\mu\tau}) \quad [5]$$

**Transcribing RNAPs:**

$$N_{\text{m}} = G_{\text{C}} N_{\text{op}} \frac{c_{\text{free}}}{c_{\text{free}} + K_{\text{m}}} \left(1 + \frac{L_{\text{m}} V_{\text{m}}}{c_{\text{m}}}\right) \quad [1a]$$

$$= \frac{r_{\text{m}}}{c_{\text{m}}} \quad [1b]$$

$$N_{\text{r}} = N_{\text{rrn}} \frac{c_{\text{free}}}{c_{\text{free}} + K_{\text{r}}} \left(1 + \frac{L_{\text{r}} V_{\text{r}}}{c_{\text{r}}}\right) \quad [2a]$$

$$= \frac{r_{\text{r}}}{c_{\text{r}}} \quad [2b]$$

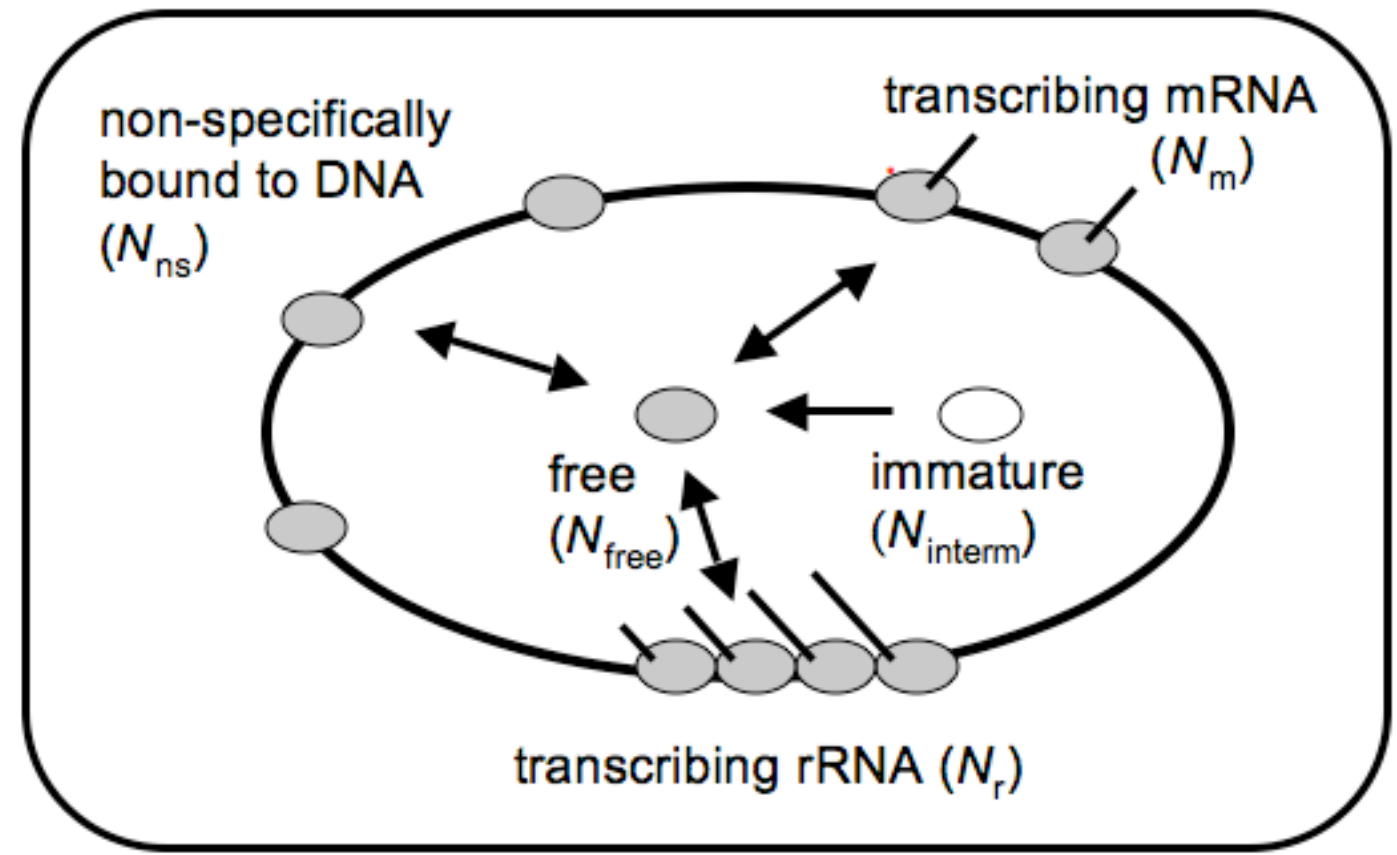

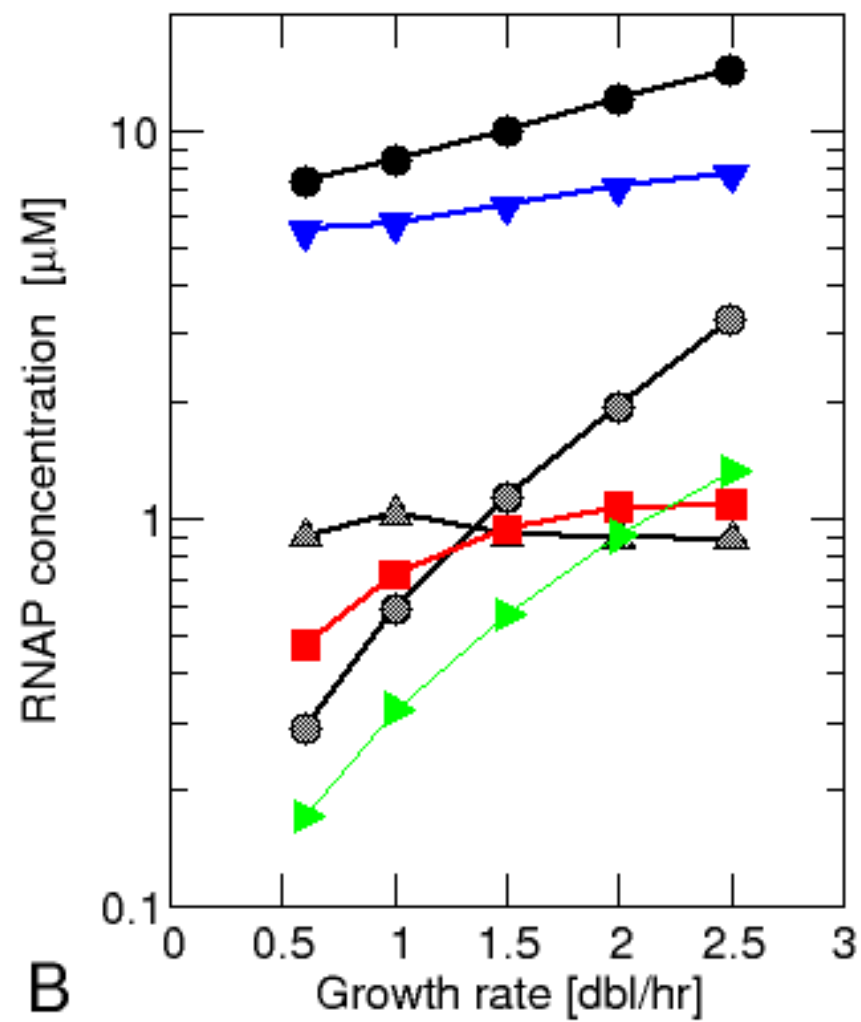

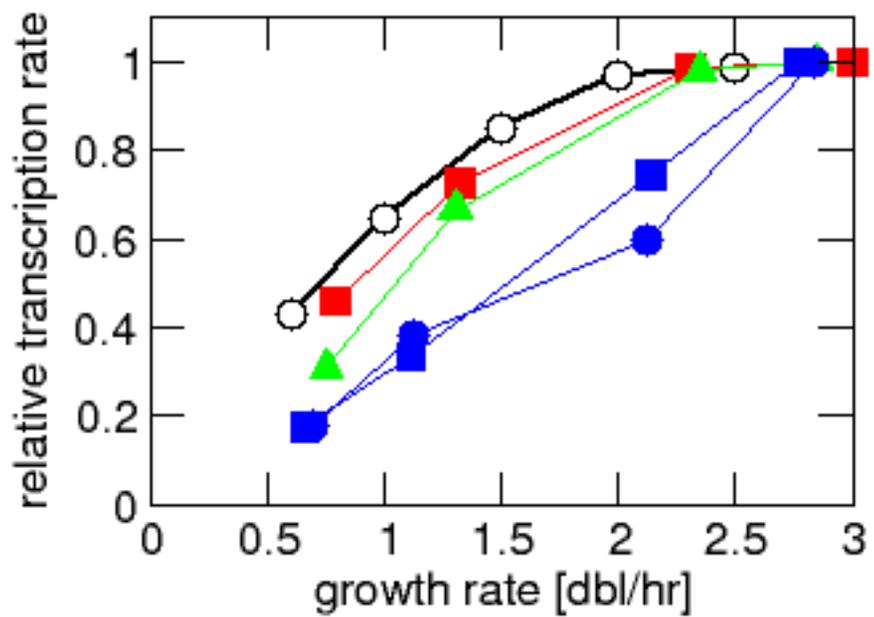

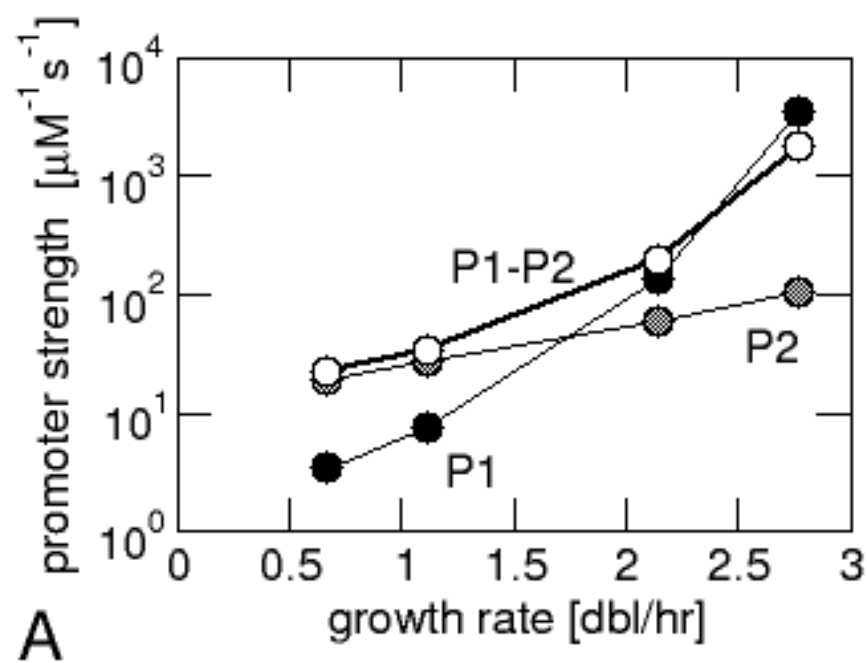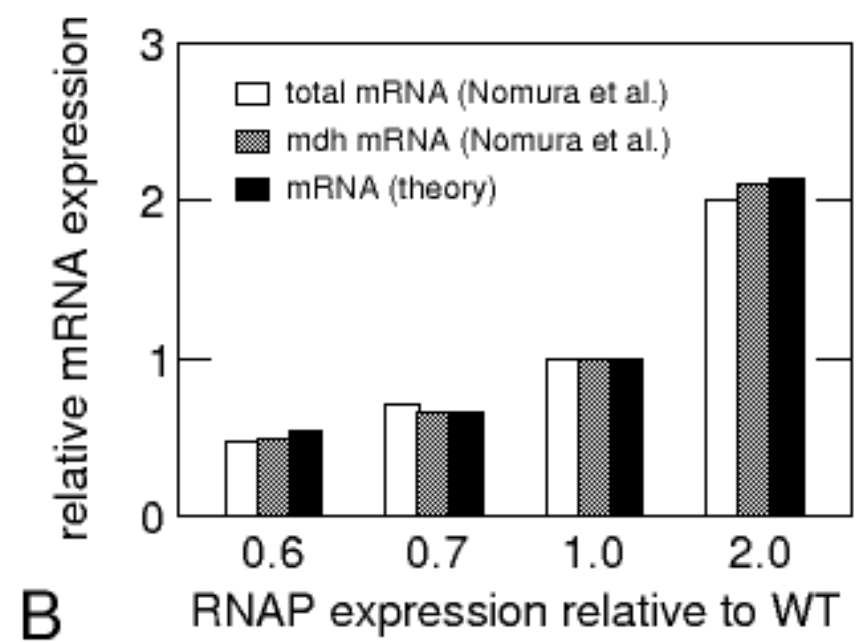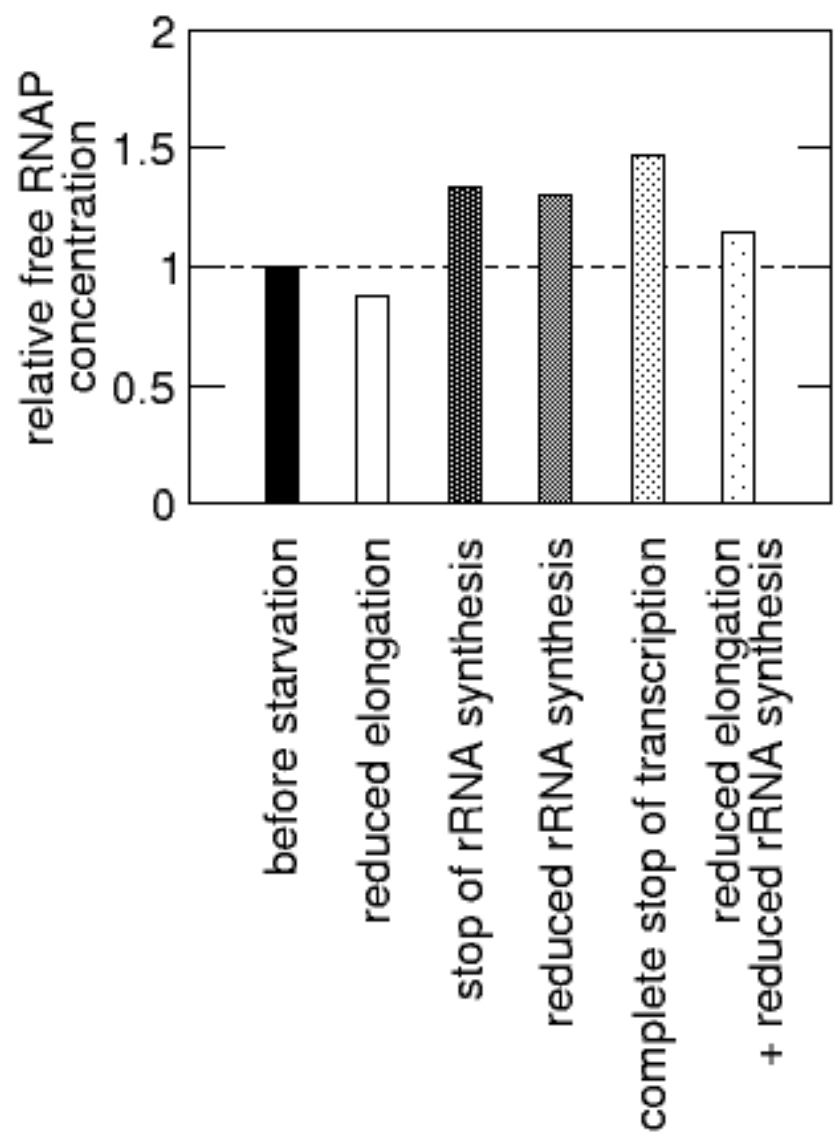

# SUPPORTING INFORMATION

# Growth-rate dependent partitioning of RNA polymerases in bacteria

Stefan Klumpp[*] and Terence Hwa

**Supporting Text:** 14 pages (including references and figure legends)

**Supporting Tables**: 2 Tables

**Supporting Figures:** 7 figures

# SUPPORTING TEXT

## METHODS

**Detailed description of the model**

To partition RNAPs into the five classes of our model, we expressed the numbers of RNAPs in each class in terms of measured microscopic quantities as follows:

To express the numbers of RNAPs transcribing mRNA and rRNA, $N_m$ and $N_r$, we use a Michaelis-Menten model for the activity of the corresponding promoters (1), an average promoter of mRNA-encoding operons and an effective promoter describing the *rrn* promoter pair P1-P2. These promoters are characterized by two parameters, their maximal transcription rates ($V_m$ and $V_r$, respectively) and their Michaelis constant ($K_m$ and $K_r$). The ratio of these two parameters ($A_m=V_m/K_m$ and $A_r=V_r/K_r$) provides a measure of the strength of the corresponding promoter (2). The transcription rate (or frequency of initiation of transcription) depends in addition on the concentration of free RNAPs, $c_{free}$, and is given by $f_m=V_m c_{free}/(K_m+c_{free})$ for each mRNA operon and likewise for rRNA. The numbers of elongating RNAPs per operon are then given by $L_m f_m/c_m$ and $L_r f_r/c_r$, where $L_m$ and $L_r$ are the length of the operons and $c_m$ and $c_r$ the transcript elongation speeds. To obtain the numbers of elongating RNAPs per cell, these numbers are multiplied by the numbers of operons per cell ($N_{rrn}$ for rRNA and $N_{op} \times G_C$ for mRNA, where $N_{op}$ is the number of different active operons on the genome and $G_C$ is the number of genome equivalents per cell). The total numbers of RNAPs involved in transcription also include promoter-bound RNAPs; the number of those is expressed as the product of the number of operons and the promoter occupation, which in the Michaelis-Menten model is given by $f_m/V_m$. In summary, we obtain the following expressions for the numbers of RNAPs transcribing mRNA and rRNA

$$N_m = G_C N_{op} f_m [1/V_m + L_m/c_m] = G_C N_{op} \frac{c_{free}}{c_{free}+K_m}[1+L_m V_m/c_m] \qquad [1a]$$

and

$$N_r = N_{rm} \frac{c_{free}}{c_{free} + K_r}[1 + L_r V_r / c_r] \qquad [2a]$$

In these expressions, the term describing promoter-bound RNAPs is usually small compared to the term describing elongating RNAPs: Even if an RNAP spends on average 50-fold more time at the promoter than at a site within the operon, as suggested by a recent estimate obtained from chromatin immunoprecipitation experiments (3), the promoter-bound RNAPs account only for ~2 percent (50/$L_m$ with $L_m$=3000) of the RNAPs involved in transcription.

A second estimate of the numbers of RNAPs transcribing mRNA and rRNA is obtained from the measured rates of overall mRNA and rRNA synthesis ($r_m$ and $r_r$) and the elongation speeds ($c_m$ and $c_r$), which leads to

$$N_m = r_m/c_m \qquad [1b]$$
$$\text{and} \quad N_r = r_r/c_r. \qquad [2b]$$

Non-specific binding of RNAPs to DNA is modeled as a binding equilibrium with dissociation constant $K_{ns}$, so that the number of non-specifically bound RNAPs, $N_{ns}$, is given by

$$N_{ns} = n_{sites} c_{free}/(c_{free} + K_{ns}), \qquad [3]$$

where $n_{sites} = g\, G_C$ is the number of binding sites, approximated by the product of genome size $g$ and number of genome equivalents per cell, $G_C$.

The number of RNAPs free in the cytoplasm is given by the concentration of free RNAPs via

$$N_{free} = c_{free} V_C \qquad [4]$$

with the cell volume $V_C$. Finally, the number of RNAP assembly intermediates is

$N_{\text{interm}} = N_{\text{RNAP}}(1-2^{-\mu\tau})$, [5]

where $\mu$ is the growth rate and $\tau$ the maturation time of newly synthesized RNAPs. The latter expression, which has also been used in Ref. (4), was derived by assuming that there is a delay of time $\tau$ after which a newly synthesized RNAP is fully assembled and functional. This assumption means that only RNAPs that had been there already a time $\tau$ earlier are functional. In exponential growth, these RNAPs are a fraction $2^{-\mu\tau}$ of the total RNAPs.

These microscopic expressions for the RNAP numbers in the five different fractions are used in three steps. We first estimated $N_m$ and $N_r$ by Eqs. [1b] and [2b] and partitioned the remaining RNAPs by solving

$N_{\text{RNAP}} - N_m - N_r = N_{\text{free}} + N_{\text{ns}} + N_{\text{interm}}$ [6]

for the concentration of free RNAPs, $c_{\text{free}}$ using parameter sets for different growth rates. All parameters needed to solve Eq. [6] are known with the exceptions of $K_{\text{ns}}$ and $\tau$, which were determined by fitting the ratio of cytoplasmic RNAP to total RNAP, $(N_{\text{free}} + N_{\text{interm}})/N_{\text{total}}$ to the minicell data (see below and Fig. S1). When these parameters are fixed, Eq. [6] leads to the predicted partitioning of RNAPs shown in Fig. 2. Finally, we used Eqs. [1a] and [2a] together with the predicted free RNAP concentration and estimates of the maximal transcription rate to determine the promoter strengths $A_r$ and $A_m$.

The resulting promoter strengths for mRNA were used to study RNAP over- and underexpression as well as the stringent response. To study RNAP over- and underexpression, we used Eq. [1a] to describe the transcription of mRNA with the determined Michaelis constant of the mRNA promoters. For the transcription of rRNA, we fixed the number of RNAPs transcribing rRNA to the value for wild-type cells with normal RNAP level to mimic the effect of feedback control. We then varied the total number of RNAPs per cell and determined the partitioning of the remaining RNAPs into the other four classes as well as the resulting transcription rate for mRNA. To study the different scenarios for the stringent response, we used the determined promoter strengths both for mRNA and rRNA, and adjusted one or several of the model parameters according to what has been measured during the stringent response.

**Parameter values**

All parameter values used in the calculations for balanced exponential growth are summarized in Tables S1 and S2. Most of these parameter values were taken from Tables 3 and 4 of the review by Bremer and Dennis (5). Parameters not given there were estimated in the following way: Numbers of RNAPs transcribing mRNA and rRNA were determined using Eq. [2]. The cell volume was calculated from the cell mass using the cell mass and volume measurements of ref. (6). The dissociation constant of non-specific binding, $K_{ns}$, and the maturation time $\tau$ of newly synthesized RNAPs were taken to be independent of growth rate and determined by fitting the fraction of RNAP that is cytoplasmic, $(N_{free}+N_{interm})/N_{RNAP}$, to the values measured using minicells at 1.23 and 2.5 doublings/hour (7, 8), as described below.

The average length of an *rrn* operon is 5400 nt according to EcoCyc (9); in our model we increased this length to 6500 nt to account for the remaining tRNA genes that are not in *rrn* operons ("appending" them to the *rrn* operons). The average mRNA transcript has a molecular weight of $10^6$ Da (10), which corresponds to an average operon length $L_m$ of ~3000 nt, an estimate consistent with an average of 2.6 genes per operon (11) and an average gene length of ~1000 nt (12). To determine the Michaelis constants, $K_m$ and $K_r$, and the promoter strengths ($A_m$ and $A_r$) of mRNA and rRNA promoters, we took the maximal transcription rates of these operons to be 90 min$^{-1}$ and 10 min$^{-1}$, respectively. These estimates are based on the highest transcription rates measured in vivo, which are in the range of 70-85 min$^{-1}$ for rRNA and 1.5-25 min$^{-1}$ for mRNA (2, 13), and the theoretically determined limits for the transcription rate [~90 min$^{-1}$ for rRNA, at most 40 min$^{-1}$ for mRNA, ref. (14)]. An estimate for the Michaelis constant $K_r$ of the *rrn* promoters during the stringent response has been obtained by extrapolating the predicted growth-rate dependence of the promoter strength (Fig. 4A) to a growth rate of zero. This leads to $K_r$~10-20 µM.

**Determination of $K_{ns}$ and $\tau$**

To determine the two unknown model parameters, the dissociation constant $K_{ns}$ for non-specific binding to DNA, and the RNAP maturation time $\tau$, we used data from the mini-cell experiments (7, 8). In these experiments, the fraction of cytoplasmic RNAPs ($N_{cyto}/N_{total}$) was measured at two different growth rates: 14% at 1.23 doubling/hr (7) and 17% at 2.5 doubling/hr (8). According to our model, the cytoplasmic RNAP consists of the free RNAP and assembly intermediates, i.e., $N_{cyto} = N_{free} + N_{interm}$. We can thus fix the two parameters by matching the fraction of cytoplasmic RNAP, i.e. ($N_{free}+N_{interm}$)/$N_{total}$, predicted according to our model with chosen values of $K_{ns}$ and $\tau$, to the above results of the mini-cell experiments (Figure S1). This procedure leads to a maturation time $\tau$ of 3.4 min and a dissociation constant for non-specific binding of 3100 μM (Fig. S1A). The maturation time $\tau$ of 3.4 min is consistent with the appearance of newly synthesized β subunit in the nucleoid after about 5 min (15), which should include maturation and transition to the nucleoid as mentioned above. Values for the dissociation constants for non-specific binding measured *in vitro* depend on ionic conditions and range between ~1 μM under low salt conditions and ~1000 μM for high salt concentration supposed to approximate physiological conditions (16) [a higher estimate for low salt has been obtained in another study (17)]. Our estimate for the *in vivo* value (3100 μM) is thus consistent with the *in vitro* results. It is also very similar to a recent *in vivo* estimate (1000 μM) of the dissociation constant for non-specific DNA-binding of the Lac Repressor (18).

**ADDITIONAL DISCUSSION**

**Growth-rate dependence of macromolecular crowding**

In our model, we have assumed that the dissociation constant for non-specific binding as well as the Michaelis constants of the promoters are not growth-rate dependent. They could however be growth-rate dependent if the degree of macromolecular crowding, i.e. the macromolecular volume fraction, is different for cells growing with different growth rates. Changes in the macromolecular volume fraction can both increase or decrease reaction rates and affinities, depending on whether the reaction is transition-state limited or diffusion-limited (19). Changes in the concentration of macromolecules have been observed for *E. coli* cells in media with increased osmolarity (20) and changed crowding has recently been proposed to play an important role in strains with reduced number of *rrn* operons (4). Direct measurements of the macromolecular volume fraction for *E. coli* are however quite limited for different growth media at fixed osmolarity. Zimmerman and Trach have measured the concentrations of macromolecules (RNA+protein) for *E. coli* grown in rich medium during exponential and stationary phase and found only a small difference between the two situations, with 0.3-0.37 g/ml in exponential growth phase and 0.34-0.4 in stationary phase (21). We expect the difference between exponential growth with different growth rates to be smaller than the difference between fast exponential growth and stationary phase, so these results suggest that macromolecular crowding should be similar at different growth rates. Furthermore very similar diffusion coefficients have been measured for the diffusion of GFP variants in the cytoplasm of cells growing in rich (22) and minimal medium (18), which also suggests that there is no big difference in crowding for different growth rates.[1]

On the other hand, the data for cell mass and volume of Ref. (6) (see also Table S1) indicate an increase of the density (mass/volume) over the range of growth rates studied here. It is possible that this increase in density is an artifact of the volume measurements. We therefore checked whether our results are changed if a constant density is used. In that case, we obtain a lower value for the maturation time τ and therefore a smaller fraction of immature RNAPs, but otherwise the results are very similar to those for a growth-rate dependent density. In particular, the growth-rate dependence of the free RNAP concentration (and thus the predicted transcription rates for constitutive promoters) is almost indistinguishable from the data shown in Fig. 3, (the absolute value of the free RNAP concentration is however slightly larger). We also obtained similar results when we used the cell volume data given by ref. (23), see also the footnote to Table S2. On the other hand, if we assume that the increase in density at faster growth is real, we can estimate an increase in the macromolecular volume fraction from ~0.24 at 0.6 doublings/hour to ~0.34 at 2.3 doublings/hour using the growth-rate dependent density (mass/volume, Table S1) and macromolecular mass fraction (5) together with the measured macromolecular volume fraction at fast growth (21). This implies that the free volume is decreased by ~13 percent at 2.4 doublings/hour compared to 0.6 doubling/hour. In this scenario, the change of macromolecular crowding would lead to an additional increase of the effective concentration of free RNAPs by ~13 percent over the range of growth rates studied here, or, equivalently, to a ~13 percent decrease of the dissociation constants for both non-specific binding to DNA and for binding to promoters. This estimate of the effect of increased crowding is small compared to the 2.3-fold increase predicted from the RNAP partitioning. We therefore expect our results to provide a very good approximation even if there is a growth-rate dependence of macromolecular crowding.

---

[1] Unfortunately the two experiments use different GFP variants, their molecular weights are however the same.

**Growth-rate dependence of the ribosomal RNA promoter P2**

As mentioned, the question whether P2 is a constitutive promoter is controversial in the literature. We therefore include a brief review of the experimental evidence for and against constitutive expression from P2 along with some comments. The claim that P2 is constitutive is mainly based on the following observations: (i) At slow growth, transcription from P2 has the same growth-rate dependence as transcription from other constitutive promoters (2). (ii) For any given growth rate, transcription from P2 is the same in strains with and without ppGpp and/or Fis, two regulators of the P1 promoter (13). While these experiments do clearly rule out strong regulation of P2, they are consistent with weak regulation of P2, in particular since both the co-variation of P2 with other constitutive promoters at low growth rates and the unchanged transcription activity of P2 in strains lacking Fis and/or ppGpp are only approximate [see Fig S3 A and B and Figs. 1 and 2 of ref. (13)]. Furthermore, the data for the growth-rate dependence of transcription from constitutive promoters (2) is only consistent with a constitutive P2 if the other promoters become saturated with RNAPs at high growth rates and not if the growth-rate dependent free RNAP concentration follows the relation predicted by our model as shown in Fig. 3.

Murray et al. (24, 25), on the other hand, have presented data in support of the claim that P2 is regulated in a growth-rate dependent way. (i) *In vitro*, ppGpp decreased the transcription rate from P2 about two-fold. Furthermore, ppGpp destabilized the open complex of promoter-bound RNAP (24, 25). It is however not obvious that these *in vitro* experiments are representative of the *in vivo* situation. (ii) Beta-galactosidase activity exhibits a pronounced growth-rate dependence when LacZ is expressed from a P2 promoter (24) (see also Fig. S3 C). This result is hard to interpret as enzyme activity under different growth conditions does not directly reflect the transcription rate, by may be also changed by a number of indirect effects such as the availability of ribosomes (which may affect the rate of translation initiation) and increased dilution of the protein due to faster growth. One can however compare the LacZ expression of different promoters under the same growth conditions as done in ref. (13). A comparison of the wild type P2 promoter with a P2 mutants [Fig. 3C and G of ref. (24), see also Fig. S3 C and D] shows that the activities of these promoters does not change in parallel over the studied range of growth rates, their ratio (mutant:wild type) decreases from 2.5 to 1.3 (Fig. S3 C). Since both promoters appear to be unsaturated with RNAPs, this means that at least one of them is regulated in some growth-rate dependent way.[2] Since *in vitro* transcription from the wild type P2 promoter is affected by ppGpp (24, 25), this result is likely to indicate a growth-rate dependent regulation of P2.

**Growth-rate dependent promoter strength for the average mRNA promoter**

To determine the strength of the average mRNA promoter $A_m$ in the same way as that of the *rrn* promoters, information about the number $N_{op}$ of operons or promoters is needed, but even without this information we can directly compute the effective promoter strength of the entire pool of mRNA promoters, $A_{mRNA}=A_m N_{op}$. We found $A_{mRNA} \approx 200$ (µM s)$^{-1}$ with little dependence on growth rate (Figure S6 B). This result has several possible interpretations: If the number of operons being transcribed at different growth rates remains rather constant, then the mRNA operons should not be strongly regulated. Alternatively, if more operons are transcribed at low growth rates (e.g., transporters and enzymes in biosynthetic pathways), then the decreases in the number of active operons at fast growth should be compensated by up-regulation of their transcription. At a growth rate of 1.5 doublings/hour, the number of different mRNA transcripts per cell is ~600 (10), yielding an estimated strength $A_m$ of ~0.3 $(\mu M\ s)^{-1}$, which is much weaker than that of the *rrn* promoter.

---

[2] Other promoter mutants studied in ref. (17) are possibly saturated with RNAP and do not yield conclusive results.

# SUPPORTING FIGURES

**Figure S1: Fit to cytoplasmic RNAP fraction from minicell data.** The fraction of cytoplasmic RNAPs, i.e. free RNAPs and assembly intermediates, was determined for different choices of two unknown parameters of our model, the dissociation constant $K_{ns}$ for non-specific RNAP-DNA binding and the RNAP maturation time τ. (A) shows parameter combinations that match the cytoplasmic RNAP fraction at growth rate of 1.23 doublings/hour [14% (7), black circles] and 2.5 doublings per hour [17% (8), triangles]. The intersection of the two curves determines the parameters $K_{ns}$ and τ. (B) Predicted growth-rate dependence of the fraction of cytoplasmic RNAPs (line and open circles), together with the experimental data from refs. (7, 8) (filled circles).

**Figure S2: Effect of the parameters τ and $K_{ns}$ on the predicted free RNAP concentration.** (A) Increasing or decreasing the RNAP maturation time τ by 50% compared to the predicted value of 3.4 s has a very small effect on the predicted free RNAP concentration $c_{free}$. (B) Increasing or decreasing the dissociation constant $K_{ns}$ for non-specific RNAP-DNA binding approximately rescales the free RNAP concentration in a linear fashion.

**Figure S3: Growth-rate dependent partitioning of RNAPs**. Fractions of the total number of RNAP in the different classes as functions of growth rates.

**Figure S4: Measurements of growth-rate dependent promoter activities:** (A) Promoter activities for promoters believed to be constitutive as reported in ref. (2). These promoters are the ribosomal protein promoter $P_{spc}$, the plasmid promoters $P_{bla}$ and $P_{RNAI}$, the promoter $P_L$ from phage λ and the *rrn* promoter P2. (B) The same data normalized to the value at the lowest

growth rate. (C) Beta-galactosidase activity obtained with LacZ expressed from the wild-type (-112 to +7) rrn P2 promoter (filled symbols) and a P2 mutant (insertion of C at -15) from ref. (24). (D) The same data normalized to the value at the lowest growth rate.

**Figure S5: Growth-rates dependence of transcription rates in wild-type and ppGpp-less cells.** Filled symbols show transcription rates for the wild-type and open symbols those for a relaxed strain (ΔrelA ΔspoT) as measured by Liang et al. (2) (A) Transcription rates for the *rrn* promoter P2, taken from Figs. 3b and 3f of ref.(2), (B) Transcription rates for the constitutive promoters $P_{spc}$, $P_{RNAI}$, $P_L$, and $P_{bla}$, taken from Figs. 2a and 2b of ref. (2).

**Figure S6: Growth-rate dependence of promoter strengths:** (A) Promoter strength $A_r$ of the *rrn* P1-P2 promoter pair as obtained from the transcription rates given in ref. (5) (dashed line) and ref. (13) (solid line). (B) Effective promoter strength $A_{mRNA}$ of the total pool of mRNA promoters.

**Figure S7: Dependence of changes during the stringent response on the growth-rate before starvation:** (A) the free RNAP concentration during the stringent response relative to the concentration before starvation, (B) the relative transcription rate of rRNA. Note that in the case of reduced mRNA elongation (red), the reduction of rRNA synthesis is a consequence of the reduced free RNAP concentration, while in the other cases, the increase of the free RNAP concentration is a consequence of the reduction of rRNA transcription. The last scenario (violet) combines both effects, but shows that the reduction of rRNA synthesis dominates.

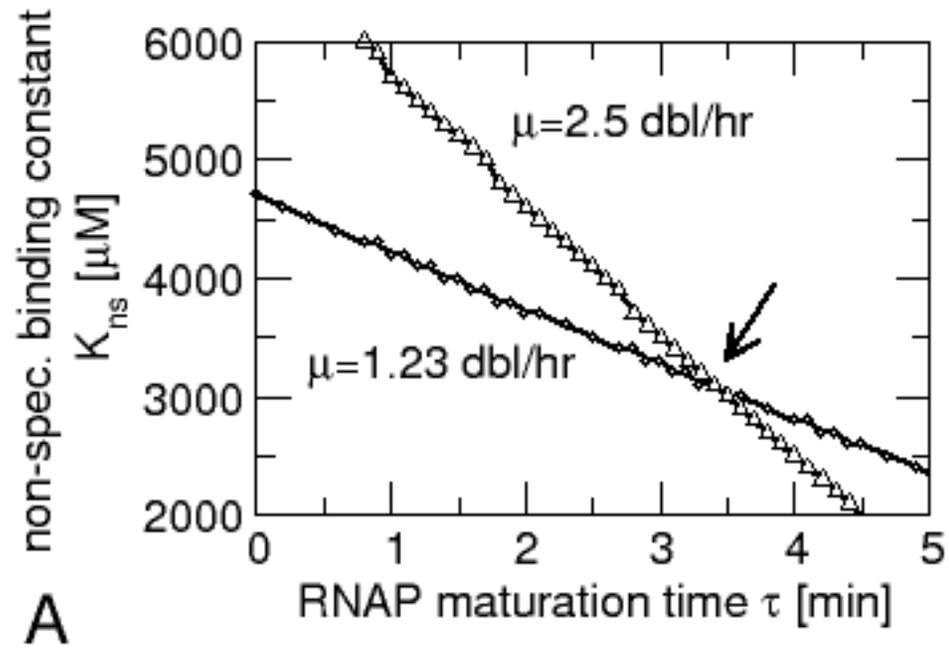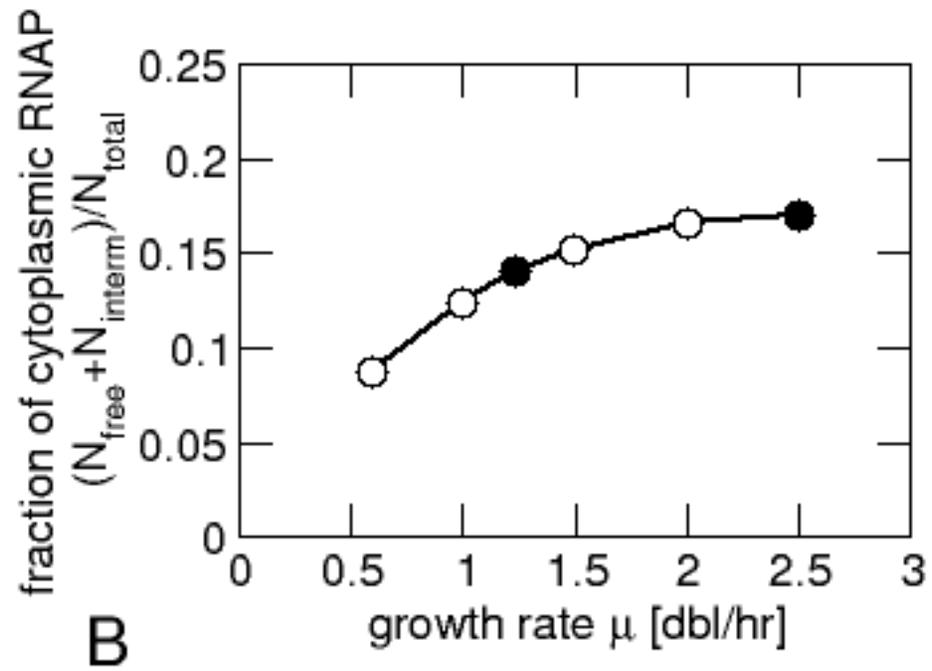

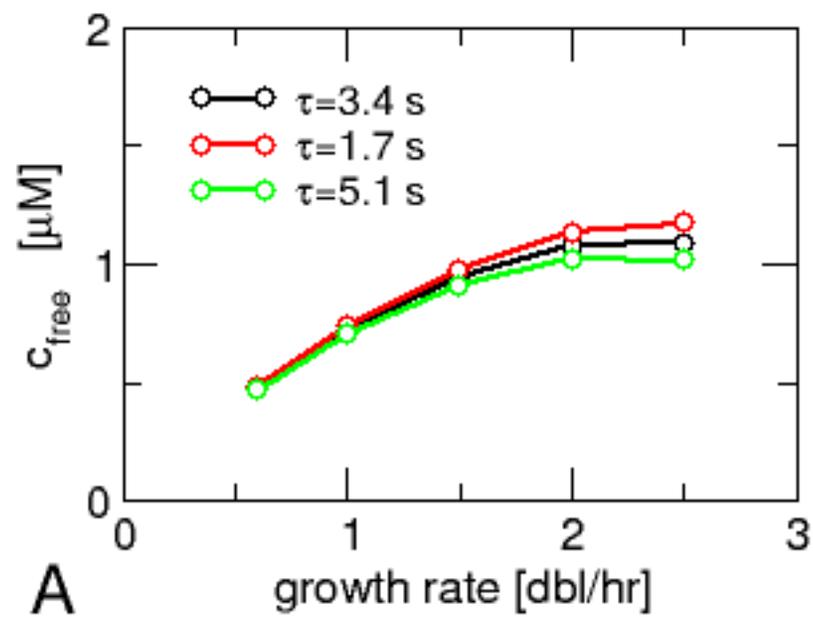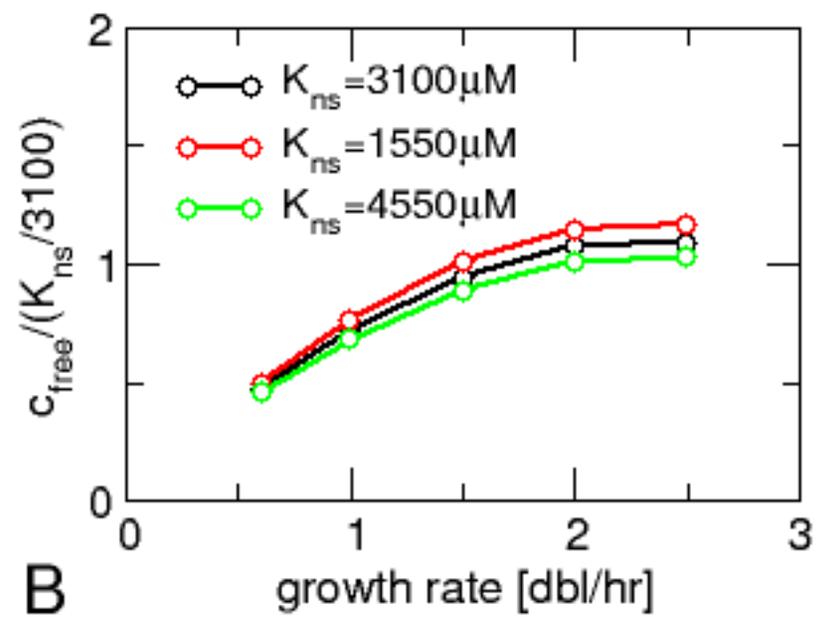

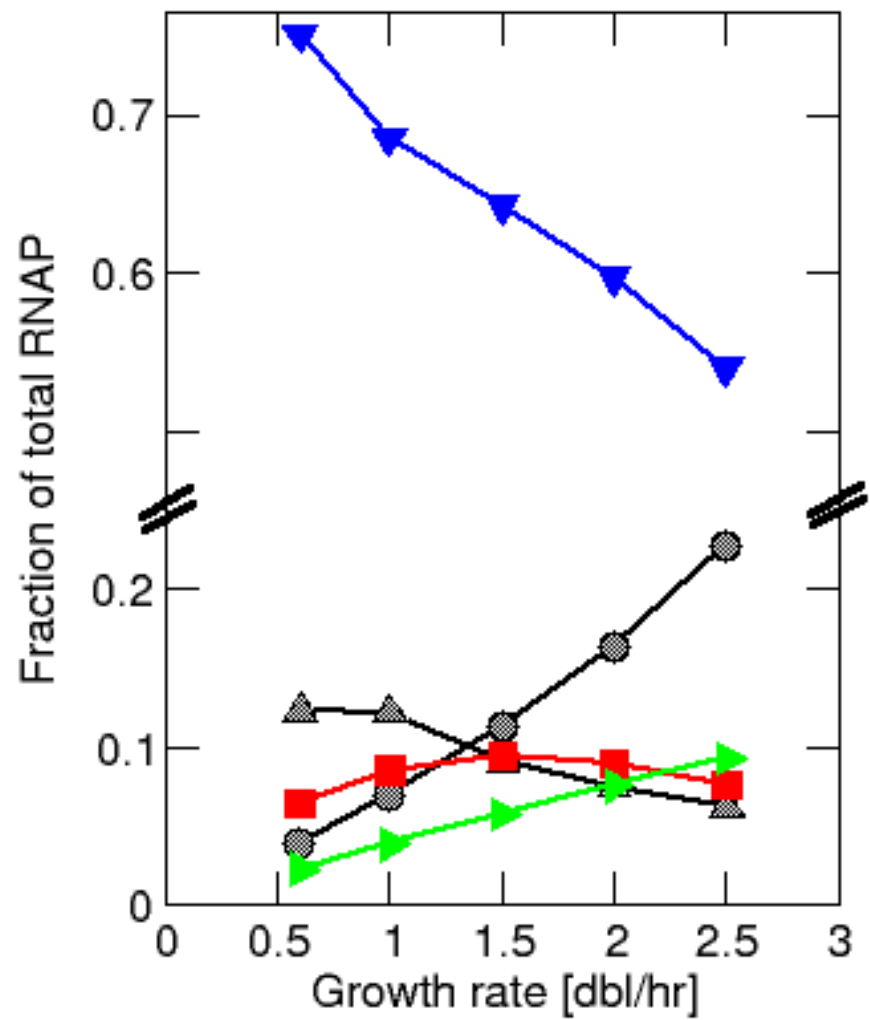

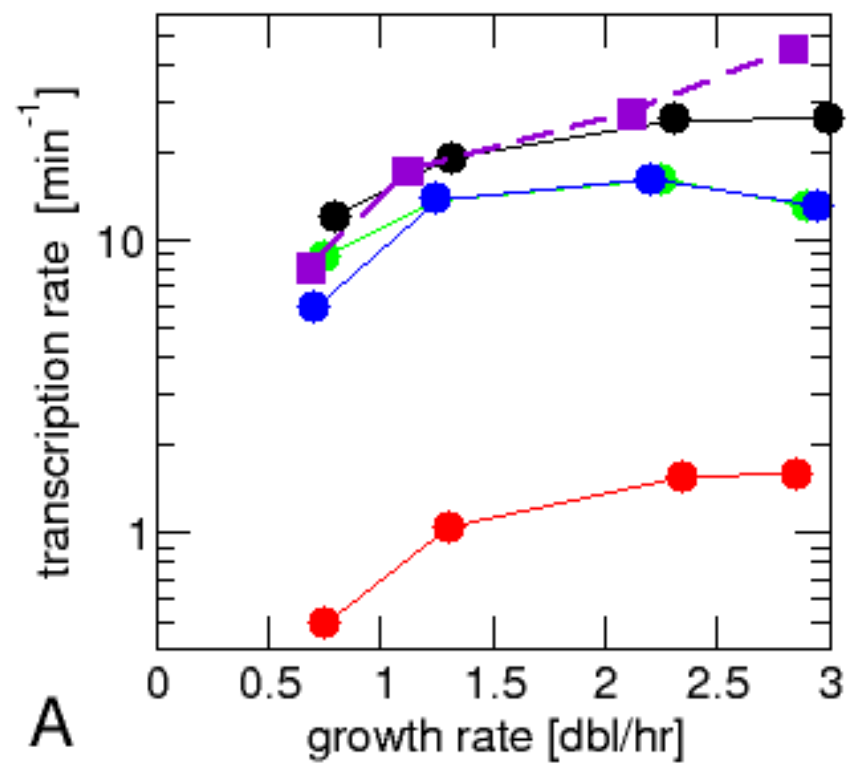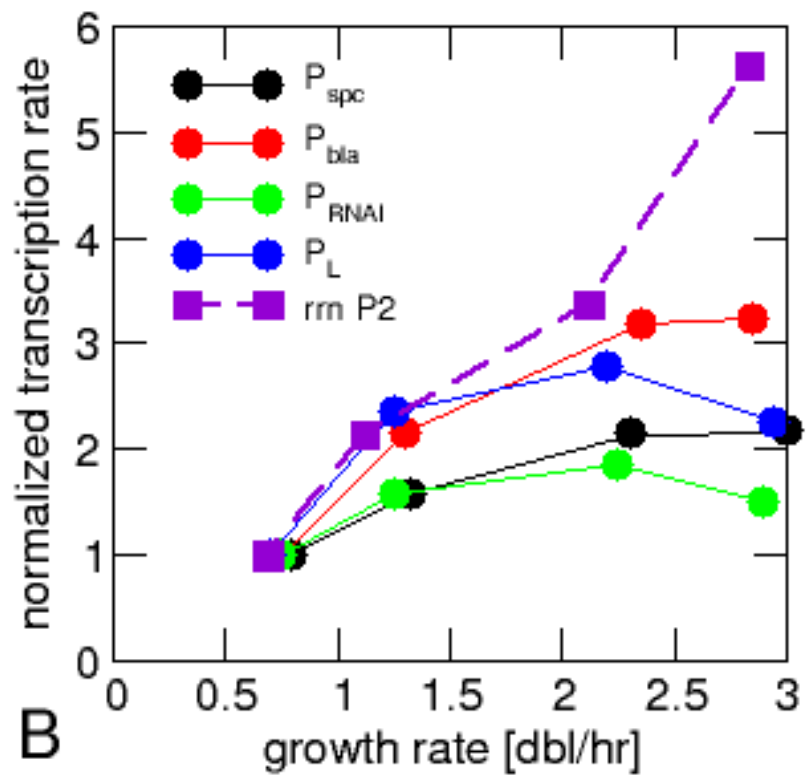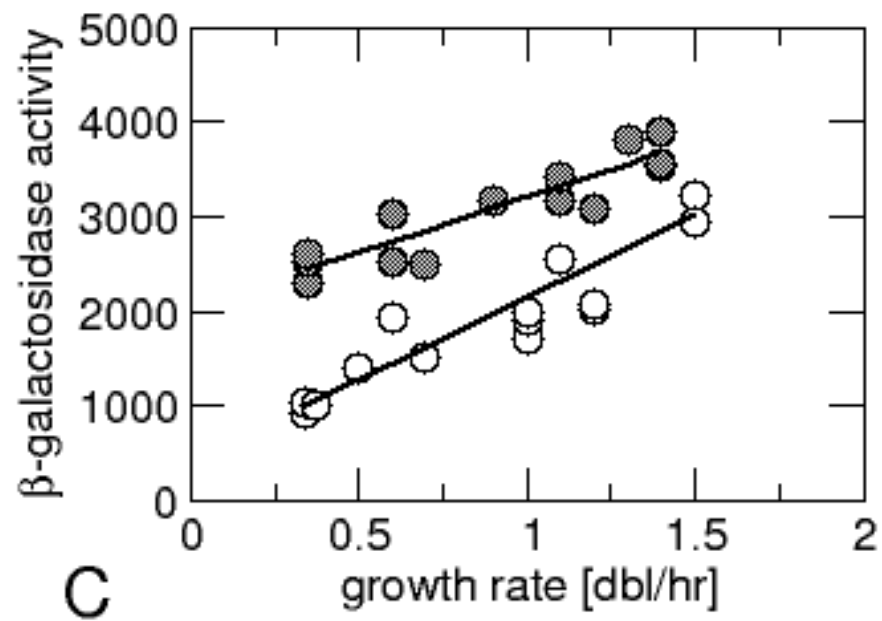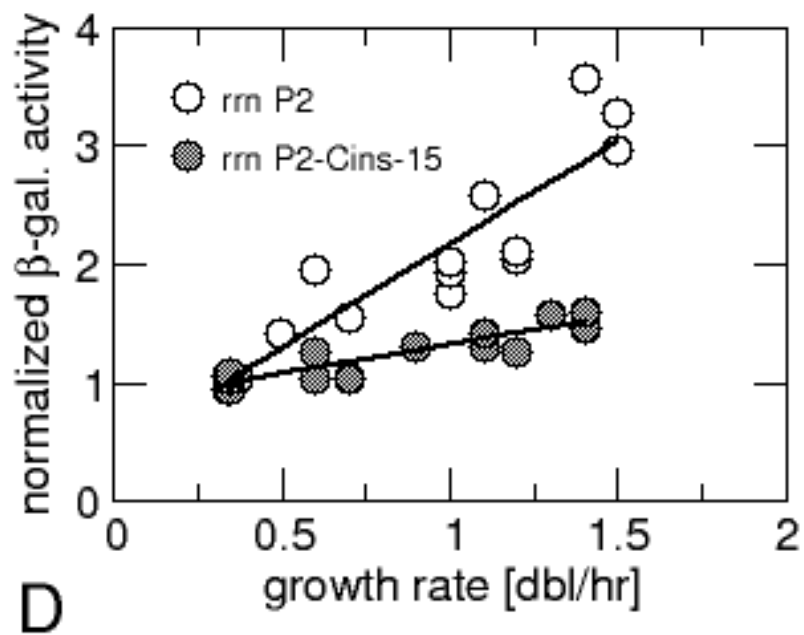

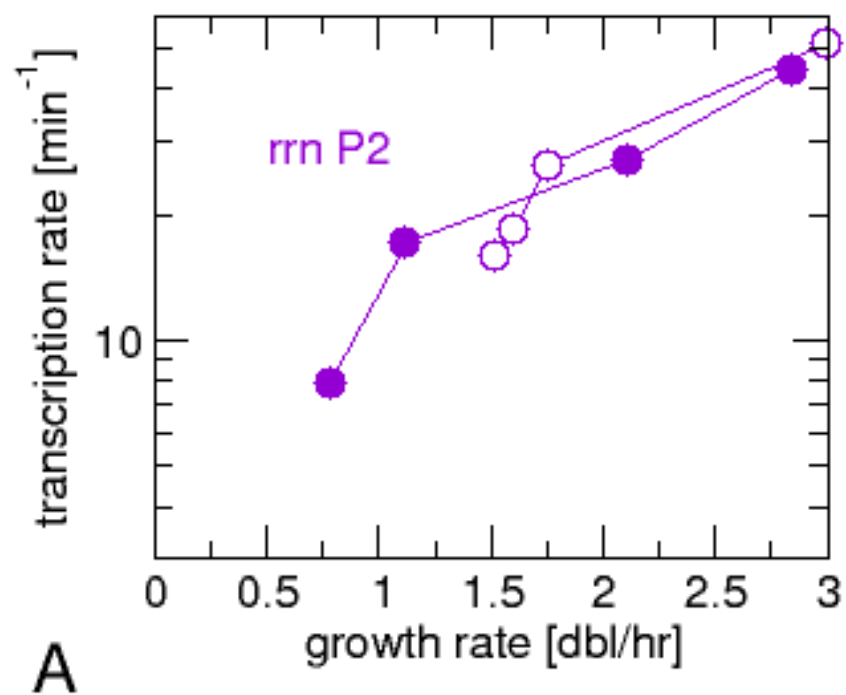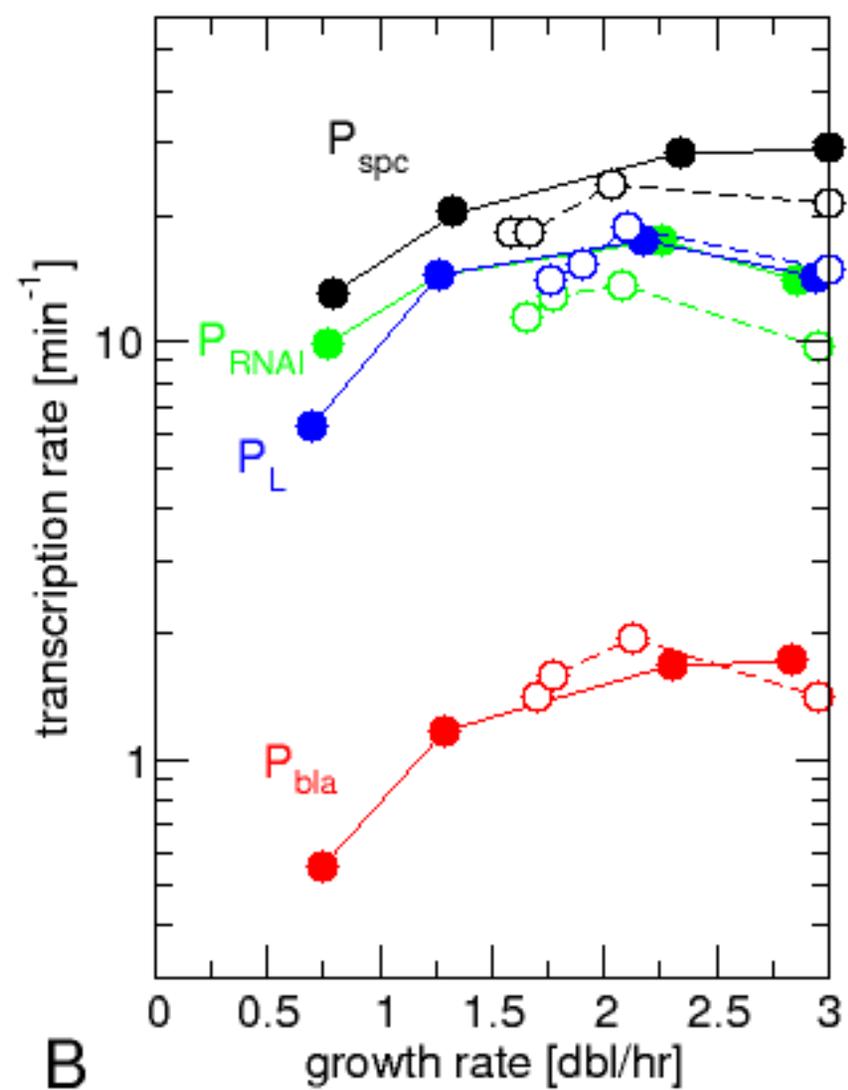

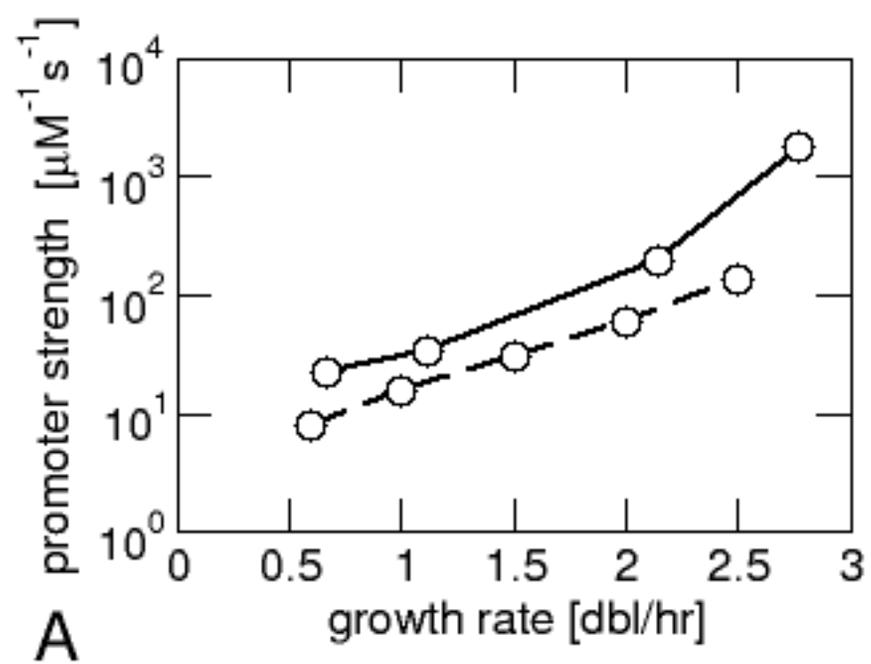 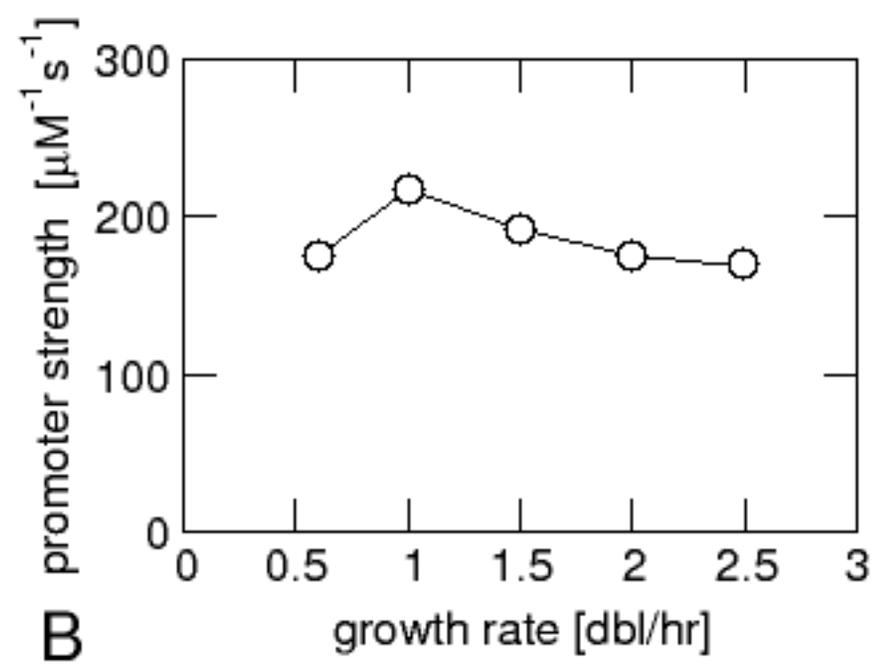

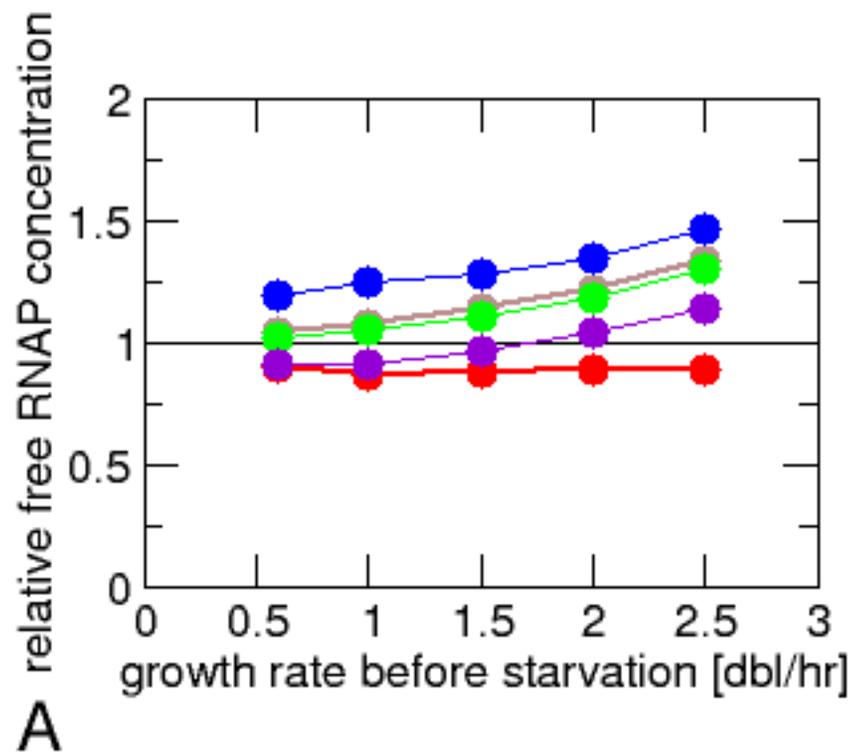
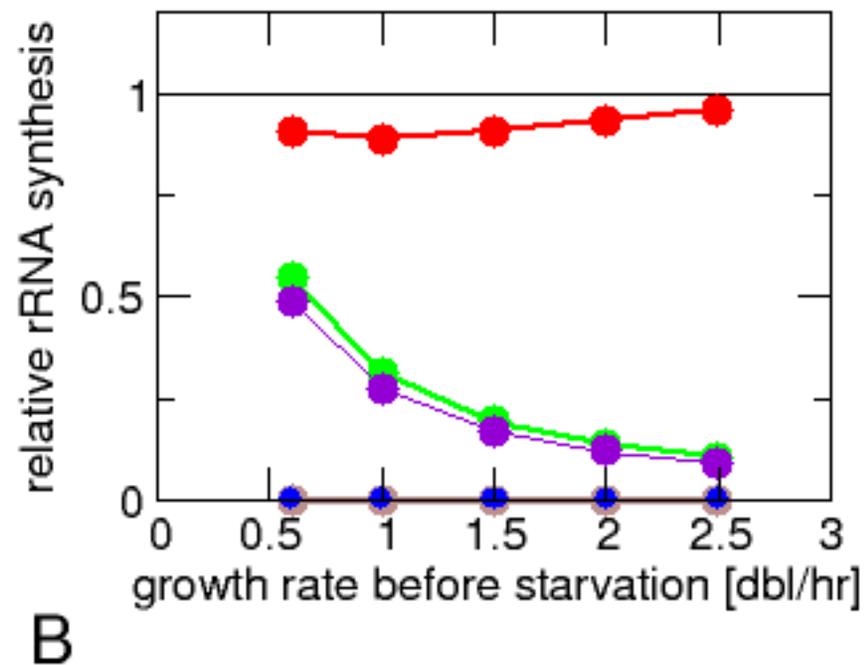

A

B

- reduced elongation
- stop of rRNA synthesis
- reduced rRNA synthesis
- complete stop of transcription
- reduced elongation + reduced rRNA synthesis

**Table S1: Growth-rate dependent parameters**

| Parameter | Symbol | Growth rate μ [dbl/hr] | | | | | Notes and references |
|---|---|---|---|---|---|---|---|
| | | 0.6 | 1.0 | 1.5 | 2.0 | 2.5 | |
| Total number of RNAP molecules per cell | $N_{total}$ | 1500 | 2800 | 5000 | 8000 | 11400 | (5) [1] |
| DNA per cell [genome equivalents] | $G_C$ | 1.6 | 1.8 | 2.3 | 3.0 | 3.8 | (5) |
| *rrn* operons per cell | $N_{rrn}$ | 12.4 | 15.1 | 20.0 | 26.9 | 35.9 | (5) |
| Mass per cell [$OD_{460}$ units/$10^9$ cells] | $M_C$ | 0.85 | 1.49 | 2.5 | 3.7 | 5.0 | (5) |
| Cell volume [μm³] | $V_C$ | 0.34 | 0.55 | 0.84 | 1.11 | 1.32 | Calculated from $M_C$ and the volumes measured in ref. (6) [2] |
| mRNA elongation speed [nt/s] | $c_m$ | 39 | 45 | 50 | 52 | 55 | (5) |
| rRNA elongation speed [nt/s] | $c_r$ | 85 | 85 | 85 | 85 | 85 | (5) |
| mRNA synthesis rate per cell [$10^5$ nt/min] | $r_m$ | 4.3 | 9.2 | 13.7 | 18.7 | 23.4 | (5) |
| rRNA synthesis rate per cell [$10^5$ nt/min] | $r_r$ | 3.0 | 9.9 | 29.0 | 66.4 | 132.5 | (5) |

| Number of RNAPs transcribing mRNA per cell | $N_m$ | 184 | 341 | 457 | 599 | 709 | calculated as $r_m/c_m$ |
|---|---|---|---|---|---|---|---|
| Number of RNAPs transcribing rRNA per cell | $N_r$ | 59 | 194 | 569 | 1302 | 2598 | calculated as $r_r/c_r$ |

[1] The numbers of total RNAPs per cell at different growth rates as given in ref. (5) and as used here are based on measurements from ref. (26), which are in good agreement with corresponding measurements from several other labs (27-30). A recent study has however reported considerably higher numbers of RNAPs per cell (31). All these studies are based on measurements of the mass fraction of total protein that is RNAP, usually called $\alpha_P$, from which the number of RNAPs per cell is obtained by multiplication with mass per cell [more precisely, these experiments determine the amounts of the β and β' subunits of RNAP, as the α subunit is known to be present in excess (27, 29, 30)]. Comparison of the measured $\alpha_P$ values shows that all studies including ref. (31) agree on the growth-rate dependence of this value and that the discrepancy between ref. (31) and the older studies is due to a unusually large amount of total protein per cell in ref. (31), about 3-fold larger than in the older studies.

[2] In ref. (6), the cell mass and volume was measured for growth rates of 1.3 dbl/hr and 2.14 dbl/hr, from these measurements, the mass per volume appears to increase slightly with growth rate, taken into account here by inter-/extrapolation. Larger values (about 1.5-fold) for the cell volume are given in ref. (23). We have also used these larger values in our calculation, and obtained very similar results (data not shown). In particular we obtained almost the same prediction for the concentration of free RNAPs (which, in the larger volume, however corresponds to a larger number of free RNAPs) and for the non-specific dissociation constant, but a smaller maturation time (1.9 min), and thus a smaller number of immature RNAPs per cell.

**Table S2: Growth-rate independent parameters**

| Parameter | Symbol | Value | Notes and references |
|---|---|---|---|
| Length of mRNA operon [nt] | $L_m$ | 3000 | see Methods |
| Length of rRNA operon [nt] | $L_r$ | 6500 | includes all tRNA genes, see Methods |
| Number of non-specific binding sites per genome | $g$ | $4.6 \times 10^6$ | from EcoCyc (9) |
| Dissociation constant for non-specific binding [µM] | $K_{ns}$ | 3100 | from fit of model to minicell data, see Methods |
| RNAP maturation time [min] | $\tau$ | 3.4 | from fit of model to minicell data, see Methods |